\documentclass{iau}

\usepackage{amsmath}
\usepackage{graphicx}
\usepackage{multirow}

\begin{document}

\lefttitle{R. Bu\v{c}\'ik, S. T. Hart, M. A. Dayeh, M. I. Desai, G. M. Mason, M. E. Wiedenbeck}
\righttitle{Origin of $^3$He abundance enhancements in gradual solar energetic particle events}

\jnlPage{1}{7}
\jnlDoiYr{2024}
\doival{10.1017/xxxxx}
\volno{388}
\pubYr{2024}
\journaltitle{Solar and Stellar Coronal Mass Ejections}

\aopheadtitle{Proceedings of the IAU Symposium}
\editors{N. Gopalswamy,  O. Malandraki, A. Vidotto \&  W. Manchester, eds.}

\title{Origin of $^3$He abundance enhancements in gradual solar energetic particle events}

\author{Radoslav Bu\v{c}\'ik$^1$, Samuel T. Hart$^{2,1}$, Maher A. Dayeh$^{1,2}$, Mihir I. Desai$^{1,2}$, Glenn M. Mason$^{3}$ and Mark E. Wiedenbeck$^{4}$}
\affiliation{$^1$Southwest Research Institute, San Antonio, TX, USA \\  $^2$University of Texas at San Antonio, San Antonio, TX, USA \\ $^3$Johns Hopkins University, Applied Physics Laboratory, Laurel, MD, USA \\  $^4$Jet Propulsion Laboratory, California Institute of Technology, Pasadena, CA, USA}

\begin{abstract}
We examined the origin of $^3$He abundance enhancement in 23 high-energy (25--50\,MeV) solar proton events that coincide with $^3$He-rich periods detected by ACE ULEIS in 1997--2021. In seven events, $^3$He enhancement was due to $^3$He leftover from preceding events or independent $^3$He events occurring during proton events. One event is the most likely impulsive ($^3$He-rich), and another is unclear. Reaccelerated remnant flare material was the most probable cause of $^3$He enhancements in the remaining 14 proton events. Imaging observations showed coronal jets in the parent active regions in six of these 14 events. Remarkably, the highest $^3$He/$^4$He occurred in events with jets, implying their contribution to $^3$He enhancement.
%We found a moderate positive correlation between $^3$He/$^4$He and Fe/O. It may suggest a common mechanism - reacceleration by coronal mass ejection-driven shock with jets contributing to the $^3$He enhancement. 
\end{abstract}


\maketitle

\section{Introduction}

Coronal mass ejection (CME)-shock-related (or gradual) solar energetic particle (SEP) events show extensive event-to-event compositional variations whose origin is largely unexplored. The events often exhibit high abundance enhancement of $^3$He and heavy ions, typical of flare-related (or impulsive) SEP events. \citet{1999ApJ...525L.133M} examined 12 gradual SEP (GSEP) events and found in eight events the 0.5--2.0\,MeV/nuc $^3$He/$^4$He between 5 and 135 times the average slow solar wind (SSW) ratio \citep{1998SSRv...84..275G}. \citet{2016ApJ...816...68D} found in 27 out of 46 GSEP events the 0.5--2.0\,MeV/nuc $^3$He/$^4$He between 2 and 194 times the average SSW ratio. Two competing ideas have been suggested: 1) reacceleration of remnant flare suprathermal ions by CME-driven shocks in corona and interplanetary (IP) space \citep[e.g.,][]{1999ApJ...525L.133M} and 2) simultaneous flare (jet)-related acceleration in parent active regions (ARs) or other sites in the corona \citep[e.g.,][]{2003GeoRL..30.8017C}.

\section{Results}

\subsection{Event selection}

We browsed \citet{2022ApJS..263...22H} catalog of $^3$He-rich periods, covering 1997 Sep 29 -- 2021 Mar 1, for coincidences with GSEP events measured by SOHO ERNE \citep{1995SoPh..162..505T}. During the SDO era (after May 2010), our interval of interest, 174 ACE ULEIS \citep{1998SSRv...86..409M} $^3$He-rich periods were reported in the catalog. We selected proton events that are sufficiently intense at high energies. Specifically, we require that the ratio between 25--50\,MeV proton intensity peak and 24-hr background before the event onset is greater than two. In addition to $^3$He periods that occurred during GSEP events (i.e., coincidences), we found cases when $^3$He periods closely ended before the events. Since these periods would contain suprathermal seed particles for further re-acceleration by CME-driven shocks, we included them in this study. Namely, we adopted $^3$He periods ending within one day before proton event onset. We note we did not find in the catalog a $^3$He-period ending within two days before proton event onset. 41 coincidences fulfill the above criteria. Ten events at low ($<$1\,MeV/nuc) ULEIS energies did not show $^4$He (and often O) intensity increase above the previous background and were not included. Furthermore, seven ion enhancements were measured during corotating interaction regions (CIRs) and one during an IP shock. These events were also excluded. The CIRs were identified based on typical solar wind plasma and IP magnetic field behavior. The IP shock is reported in the database by \citet{2023FrASS..1040323O}. Thus, we have 23 events left to explore.

\begin{table}
 \centering
 \caption{25--50\,MeV proton event characteristics}\label{table1}
 {\tablefont\begin{tabular}{@{\extracolsep{\fill}}cccccccccc}
    \midrule
    \#& \#&
     Proton event start day&
     \multicolumn{2}{c}{GOES X-ray flare}&
      \multicolumn{2}{c}{CME}& Jet& $^3$He/$^4$He ($\times$10$^{2}$)&
     Ref.\\
     \cline{4-7}
     &&&Start& Class & Speed & Width&&&\\
    \midrule
1 & 470 & 2010-Jun-12$^\text{b}$& 00:54:00& M2.0& 486	&119&N&	1.54$\pm$0.18&i, j, k, l\\
2 &473 &	2010-Aug-31$^\text{a,b}$&	…&	…& 1304&360&	Y&	0.68$\pm$0.39&	i, k, l\\
3 &	494&	2011-Jan-28$^\text{b}$&	00:44:00&	M1.3&	606&	119&	Y&	4.08$\pm$0.93&	i\\
4&	508	&2011-Apr-21$^\text{a,b}$&	…	&…	&475&	111&	Y&	43.57$\pm$19.83&	i\\
5&	524&	2011-Aug-02$^\text{b}$	&05:19:00&	M1.4&	712&	268&	N&	…	&i\\
6&	525&	2011-Aug-08&	18:00:00&	M3.5&	1343&	237&	N&	1.05$\pm$0.09&	i\\
7&	539&	2011-Nov-26$^\text{b}$	&06:09:00&	C1.2&	933&	360&	N&	0.32$\pm$0.03&	i, m\\
8&	554	&2012-Jan-19$^\text{b}$	&13:44:00&	M3.2&	1120&	360&	N	&0.99$\pm$0.08&	i\\
9&	566&	2012-May-17$^\text{b}$&	01:25:00&	M5.1&	1582	&360&	N	&0.35$\pm$0.04&	i, m\\
10&	571&	2012-Jun-14$^\text{b}$&	12:52:00&	M1.9&	987&	360&	N&	0.29$\pm$0.06&	i\\
11&	604&	2013-Feb-26$^\text{a,b}$&	…	&…	&987&	360&	N&	0.59$\pm$0.19&	i\\
12&	611&	2013-May-02&	04:58:00&	M1.1&	671&	99&	Y	&20.05$\pm$2.57&	i, j, n\\
13&	641&	2013-Sep-30$^\text{b}$&	21:43:00&	C1.2&	1179&	360&	N&	0.36$\pm$0.04	&i, j\\
14&	647&	2013-Oct-25$^\text{b}$& 07:53:00&	X1.7&	587&	360&	Y&	2.45$\pm$0.33&	i\\
  &      &                   & 14:51:00&	X2.1&	1081&360&	N&&	i\\			
15& 	665	& 2013-Dec-26$^\text{a,b}$& 	…	& …	& 1011& 	$>$171& 	N& 	2.28$\pm$0.56	& i\\
16& 	668	& 2014-Jan-04& 	18:47:00& 	M4.0& 	977& 	360& 	Y& 	0.29$\pm$0.08& 	m\\
17& 	682	& 2014-Feb-25$^\text{b}$& 	00:39:00& 	X4.9& 	2147& 	360& 	N	& 0.30$\pm$0.03& 	\\
18& 	692	& 2014-Apr-18& 	12:31:00& 	M7.3& 	1202& 	360& 	Y& 	0.69$\pm$0.06& 	j\\
19& 	700& 	2014-May-07	& 16:07:00& 	M1.2& 	923& 	360& 	Y& 	1.96$\pm$0.57& 	\\
20& 	715& 	2014-Jun-12	& 21:34:00	& M3.1& 	684	& 186& 	Y	& 4.20$\pm$0.30& 	\\
21& 	781	& 2015-Jul-19& 	09:22:00& 	C2.1& 	782& 	194& 	N	& 0.35$\pm$0.27& 	\\
22& 	789	& 2015-Sep-20	& 17:32:00& 	M2.1& 	1239	& 360& 	N	& 0.49$\pm$0.07	& \\
23& 	790& 	2015-Sep-30& 	16:54:00& 	C2.4& 	...	& ...	& Y	& 0.58$\pm$0.06	& j\\
    \midrule
    \end{tabular}}
\tabnote{\textit{Notes.} $^\text{a}$Events with backside source. $^\text{b}$Events measured at least on one of the two STEREO spacecraft.}
\tabnote{\textit{References.} (i) \citet{2014SoPh..289.3059R}, (j) \citet{2019ApJ...877...11C}, (k) \citet{2013ApJ...762...54W}, (l) \citet{2016ApJ...833...63B}, (m) \citet{2016ApJ...816...68D}, (n) \citet{2015ApJ...806..235N}}
\end{table}

Table~\ref{table1} shows the solar characteristics of proton events. Column 1 marks the event number. Column 2 shows the corresponding $^3$He-rich period number \citep{2022ApJS..263...22H}. Column 3 indicates the start day of the proton event at 25--50\,MeV. Columns 4 and 5 show GOES X-ray start time (with day in column 3) and class as obtained from the NOAA Space Weather Prediction Center (SWPC) Edited Events list. Events \#2, 4, 11, and 15 have a backside source. A flare for event \#13 started on an earlier day than the proton event start day. Note that event \#14 was associated with multiple flares. Columns 6 and 7 indicate CME speed (km/s) and width ($^\circ$) from the LASCO manually identified CME catalog\footnote{https://cdaw.gsfc.nasa.gov/CME$\_$list/}  \citep{2004JGRA..109.7105Y}. Event \#23 has not reported CME. Column 8 indicates with Y a jet in the parent AR as observed by SDO AIA \citep{2012SoPh..275...17L} or STEREO EUVI \citep{2008SSRv..136...67H}. Column 9 shows 0.5--2.0\,MeV/nuc $^3$He/$^4$He. Note that independent ISEP events that occurred during some proton events were excluded from the $^3$He/$^4$He calculation (details in Section~\ref{obs}). The last column marks references to earlier studies of these events. All events except \#21 and \#23 were associated with type II radio bursts. The radio bursts were reported in the SWPC list or identified using STEREO-A, -B \citep{2008SSRv..136..487B}, or Wind \citep{1995SSRv...71..231B} radio spectrograms. 14 events were measured at least on one of the two STEREO spacecraft by HET \citep{2008SSRv..136..391V} at the proton energy range 24--41\,MeV. The minimum longitudinal separation between Earth and STEREO was 70$^\circ$, implying that these events were widespread. Five events (\#7, 9, 13, 17, and 18) are on the NOAA list of Solar Proton Events Affecting the Earth’s Environment\footnote{https://umbra.nascom.nasa.gov/SEP/}.   

\subsection{Observations}~\label{obs}
  
\begin{figure}
 \centering
      \includegraphics[scale=.5,angle=90]{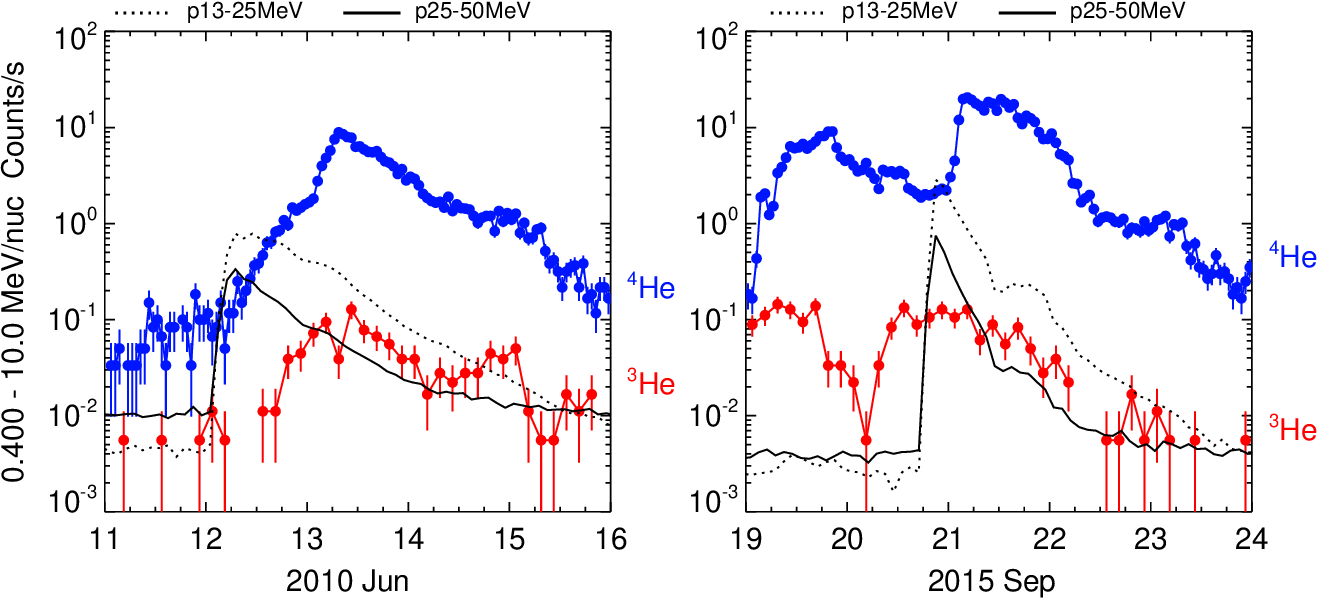}
    \caption{ACE ULEIS 3-hr $^3$He (2.8--3.1\,AMU) and 1-hr $^4$He (3.5--6.0\,AMU) count rates in events \#1 (left) and \#22 (right). Overplotted are SOHO ERNE proton fluxes (in arbitrary units) in two energy ranges.}
    \label{figure1}
  \end{figure}

It has been argued that similar time profiles of $^3$He with other ion species suggest the common acceleration and transport origin \citep{1999ApJ...525L.133M,2000AIPC..528..107W}. Four events (\#15, 18, 21, 22) show distinct time profiles of $^3$He and $^4$He, caused by $^3$He left over from preceding impulsive ($^3$He-rich) events. Six events (\#1, 3, 4, 5, 9, 14) from the remaining 19 events contain independent $^3$He-rich SEP enhancement, which also leads to distinct time profiles. The periods of independent $^3$He enhancements were excluded from four events (\#1, 5, 9, 14) by adjusting the integration interval. It was not possible in two events (\#3, 4) where the independent $^3$He-rich period spanned nearly the whole proton event. Thus, we have 17 events to explore. Figure~\ref{figure1} (left) shows event \#1 (included in the final list) with preceding $^3$He ions, similar time profiles of $^3$He and $^4$He during proton flux decay until the beginning of 2010 Jun 14. Later, the increase in $^3$He count rates is due to an independent ISEP event accompanied by solar energetic electrons detected by ACE EPAM \citep{1998SSRv...86..541G}. Figure~\ref{figure1} (right) shows event \#22 (excluded from the final list) with $^3$He left over from the preceding ISEP event. During a sudden rise in $^4$He count rates on 2015 Sep 21, associated with the GSEP event, $^3$He count rates continued to decrease.   

\begin{figure}
 \centering
      \includegraphics[scale=.48, angle=90]{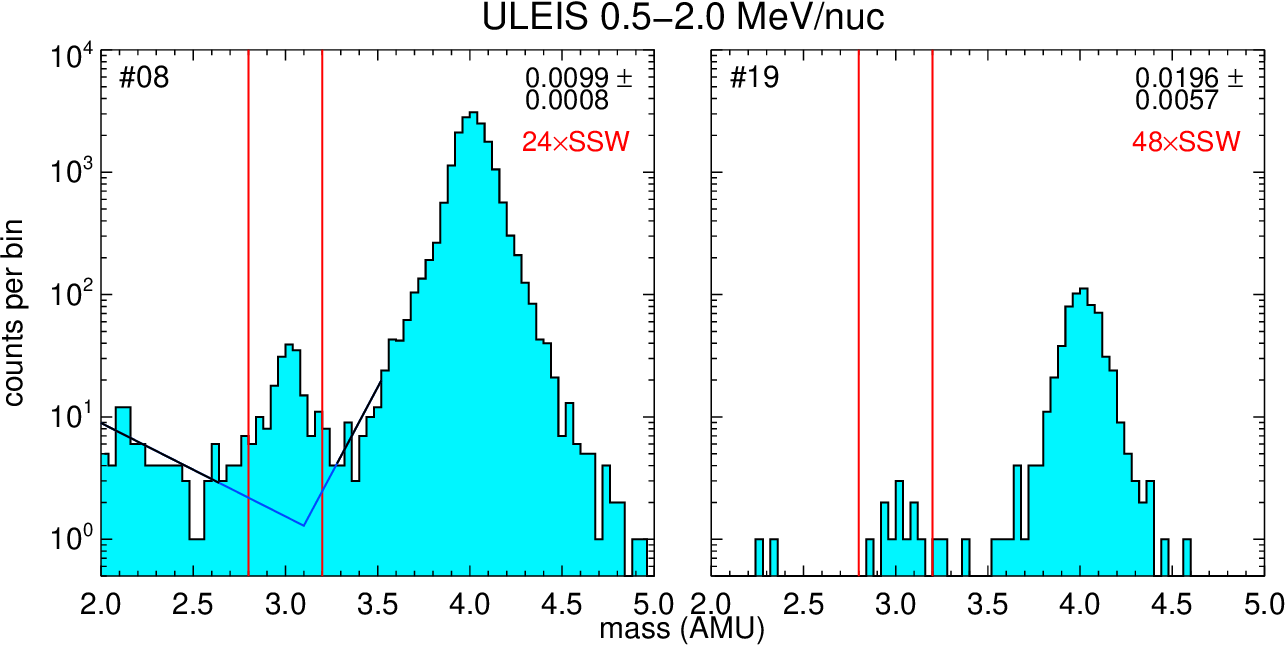}
    \caption{He mass histograms for events \#8 (left) and \#19 (right). The vertical red lines mark the $^3$He mass range. The $^3$He/$^4$He ratios are shown in the upper right corner. Left: The two tilted lines are the least square fits of the background and $^4$He spillover (black) and the extrapolations to the $^3$He range (blue).}
    \label{figure2x}
  \end{figure}

The $^3$He/$^4$He ratio was determined using helium (2--5\,AMU) mass histograms. The ratio is calculated in the broad energy range of 0.5--2.0\,MeV/nuc, ensuring enough $^3$He counts. We fitted the background at 2.00--2.64\,AMU and spillover from $^4$He at 3.28--3.52\,AMU with two $\log(y)=a + bx$ functions, where $a$ and $b$ are parameters, and $x$ corresponds to mass and $y$ to ion counts. Both fits were extrapolated to the $^3$He range (2.8--3.2\,AMU), and the number of counts under the fit was subtracted from the total number of $^3$He counts. Net $^4$He counts were obtained from the 3.5--5.0\,AMU range. Figure~\ref{figure2x} shows event \#08, where background and spillover were removed, and event \#19, with no background and spillover. After adjusting integration intervals to eliminate independent $^3$He events, the $^3$He counts in \#5 dropped to zero. In the remaining events, the $^3$He peaks are finite - the number of net $^3$He counts is 2$\sigma$ or more above zero \citep[same as in][]{2006ApJ...649..470D}. 
 
\begin{figure}
 % \centering
      \includegraphics[scale=.49]{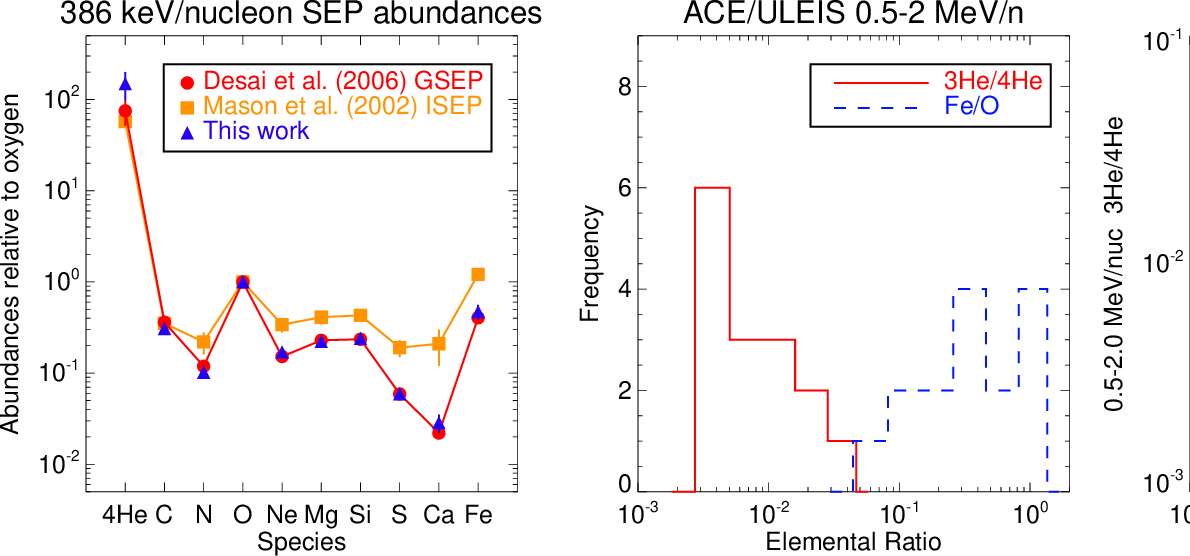}
    \caption{Left: Survey-averaged abundances relative to O in this work (blue triangles), GSEP (red circles), and ISEP (orange squares) reference abundances. Middle: Distribution of 0.5--2.0\,MeV/nuc $^3$He/$^4$He (red solid line) and 386\,keV/nuc Fe/O (blue dashed line). Right: $^3$He/$^4$He vs. Fe/O. The three numbers indicate the Spearman correlation coefficient, $p$-value, and sample size. }
    \label{figure2}
  \end{figure}

Figure~\ref{figure2} (left) displays the survey-averaged abundances using 16 events in Table~\ref{table1} (\#1, 2, 5--11, 13, 14, 16, 17, 19, 20, 23). Event \#12 was not included in the averaging. This event was on the ISEP event list by \citet{2015ApJ...806..235N}. Indeed, the event 0.5--2.0\,MeV/nuc $^3$He/$^4$He $=$ 0.20$\pm$0.03 (492$\times$SSW value) is exceptionally high. Furthermore, the abundances of Si, S, and Fe are ISEP-like, and Ne and Mg are enhanced compared to GSEP values. Figure~\ref{figure2} (left) shows that the relative abundances in this work are consistent with reference GSEP abundances. Table~\ref{table2} lists the average heavy-ion abundances of this survey along with GSEP \citep{2006ApJ...649..470D} and ISEP \citep{2002ApJ...574.1039M} average abundances. Figure~\ref{figure2} (middle) indicates that $^3$He/$^4$He peaks at the lowest values while Fe/O shows a more uniform distribution. Some high Fe/O can be due to rigidity-dependent transport effects. This will be explored in the next study. Minimum, maximum, mean, and median $^3$He/$^4$He expressed as multipliers of the SSW value (4.08$\pm$0.25)$\times$10$^{-4}$ are 7, 103, 26$\pm$7, and 15, respectively. Figure~\ref{figure2} (right) shows a moderate positive correlation ($r=$0.593) between $^3$He/$^4$He and Fe/O with a low probability of 2\% that ratios are a random sample. Note that because $^3$He dropped to zero in \#5, both the middle and right panels involve 15 events. \citet{2006ApJ...649..470D} reported $r=$0.66 (for 29 events with a finite $^3$He peak), but after excluding two outliers, $r$ dropped to 0.026. The authors concluded that Fe/O is independent of $^3$He/$^4$He in GSEP events. Similar observations have been reported in ISEP events \citep{1986ApJ...303..849M,1994ApJS...90..649R}. 

\begin{table}
 \centering
 \caption{Average heavy ion abundances}\label{table2}
 {\tablefont\begin{tabular}{@{\extracolsep{\fill}}cccc}
    \midrule
    Element & GSEP$^\text{a}$ & GSEP$^\text{b}$ & ISEP$^\text{c}$ \\
    &(386\,keV/nuc)&(385\,keV/nuc)&(385\,keV/nuc)\\
    \midrule
$^4$He & 149.5 $\pm$ 50.7 &75.0 $\pm$ 23.6 & 57 $\pm$ 7.79\\
C&	0.308 $\pm$ 0.014&	0.361 $\pm$ 0.012&	0.35 $\pm$ 0.08\\
N&	0.102 $\pm$ 0.004&	0.119 $\pm$ 0.003&	0.22 $\pm$ 0.06\\
O&	$\equiv$1.0 $\pm$ 0.02	&$\equiv$1.0 $\pm$ 0.02&	$\equiv$1.0 $\pm$ 0.14\\
Ne&	0.170 $\pm$ 0.015&	0.152 $\pm$ 0.005&	0.34 $\pm$ 0.06\\
Mg& 0.223 $\pm$ 0.018&	0.229 $\pm$ 0.007&	0.41 $\pm$ 0.07\\
Si&	0.240 $\pm$ 0.022&	0.235 $\pm$ 0.011&	0.43 $\pm$ 0.06\\
S&	0.059 $\pm$ 0.005&	0.059 $\pm$ 0.004&	0.19 $\pm$ 0.04\\
Ca&	0.029 $\pm$ 0.006&	0.022 $\pm$ 0.002&	0.21 $\pm$ 0.09\\
Fe&	0.472 $\pm$ 0.087&	0.404 $\pm$ 0.047&	1.21 $\pm$ 0.14\\
\midrule
  Ratio & (0.5--2.0\,MeV/nuc)&(0.5--2.0\,MeV/nuc) &(385\,keV/nuc) \\
  \midrule
  $^3$He/$^4$He&0.011$\pm$0.003 &0.006$\pm$0.002 &0.354$\pm$0.137 \\
    \midrule
    \end{tabular}}
    \tabnote{\textit{Notes.} $^\text{a}$This work. $^\text{b}$From \citet{2006ApJ...649..470D} . $^\text{c}$From \citet{2002ApJ...574.1039M}.}
\end{table}

\begin{figure}
\centering
  \includegraphics[scale=.6]{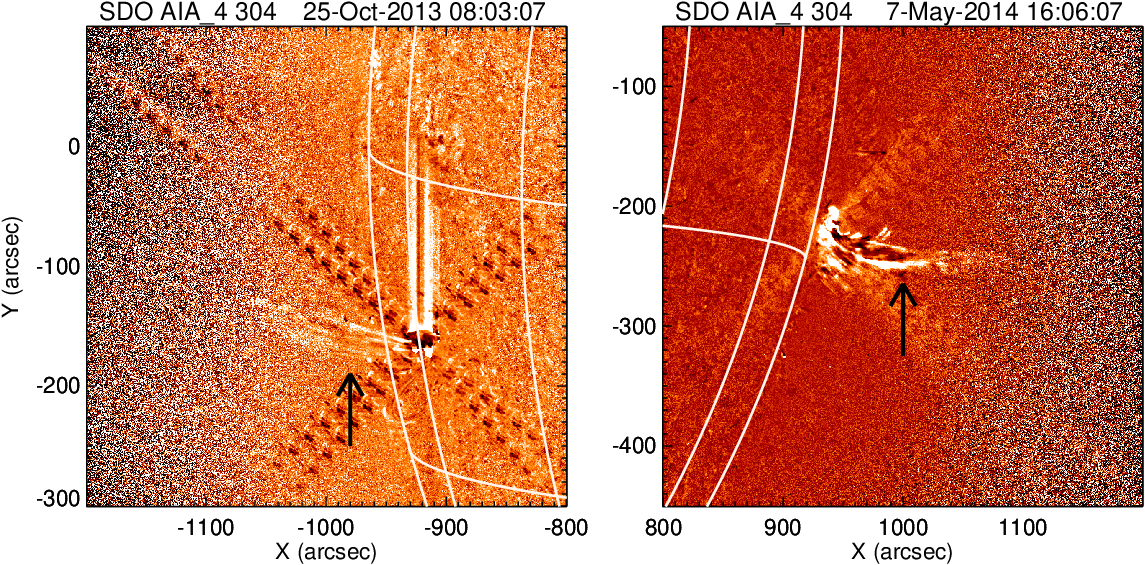}
  \caption{1-min running difference SDO AIA 304\,{\AA} images of source flares in events \#14 (Left panel; X1.7 flare at E75) and \#19 (Right panel; M1.2 flare at W90). Jet expulsions are marked by arrows. Vertical bright stripes in the left panel are saturation, and diagonal patterns in both panels are diffraction fringes due to a strong EUV signal.}
\label{figure3}
\end{figure}

Figure~\ref{figure3} shows the source flare for two events. Along with the bright flares that left some artifacts in the EUV images, ejections of jets from the sources were observed. Jets, a signature of magnetic reconnection involving field lines open to IP space, have been associated with the source of $^3$He-rich SEPs \citep[e.g.,][and references therein]{2020SSRv..216...24B,2023FrASS..1048467N}. 

Note that events \#1 and \#2 were included in the list of ISEP events by \citet{2013ApJ...762...54W} and \citet{2016ApJ...833...63B}. The events had $^3$He/$^4$He of 38$\times$SSW and 17$\times$SSW ratios, respectively. These values are typical for $^3$He-enriched GSEP events. The heavy ions showed mixed composition (N, S with GSEP and Ne, Fe with ISEP abundances) in event \#1 and GSEP composition (except Fe) in event \#2. In addition, these events were widespread and associated with type II radio bursts. 

\section{Conclusion}

We explored the $^3$He origin in 23 high-energy (25--50\,MeV) proton events. Except for independent ISEP events starting just before or during proton events, the $^3$He origin in 15 events has no simple cause (presumed ISEP event \#12 is not counted). 14 out of 15 events in this study showed preceding remnant $^3$He. All these 14 events showed similar $^3$He \& $^4$He intensity-time profiles, suggesting that reacceleration of remnant material by the associated CME shock was the most probable cause of $^3$He enhancement. The origin of $^3$He ions in one event (\#10) remains unclear. In 6 events (those with preceding remnant $^3$He ions), there were observed simultaneous jets in parent ARs. Interestingly, three events (\#19, 14, 20) with the highest $^3$He/$^4$He were associated with jets. This signifies the role of jets as a simultaneous source of energetic $^3$He in GSEP events.

\bigskip

{\small R.B. acknowledges support by NASA grants 80NSSC21K1316 and 80NSSC22K0757. The LASCO C2 CME catalog is generated and maintained at the CDAW Data Center by NASA and The Catholic University of America in cooperation with the Naval Research Laboratory. SOHO is a project of international cooperation between ESA and NASA.}

 \bibliographystyle{iaulike_my}% style aa.bst
   \bibliography{Sample_1} % your references Yourfile.bib

\end{document}